\documentclass[9pt, conference]{IEEEtran}
\IEEEoverridecommandlockouts

\usepackage{cite}
\usepackage{amsmath,amssymb,amsfonts}
\usepackage{algorithmic}
\usepackage{graphicx}
\usepackage{textcomp}
\usepackage{xcolor}

\usepackage{mathptmx} % This is Times font
\usepackage{upgreek}
\usepackage{textcomp}
\usepackage{booktabs}
\usepackage[hidelinks]{hyperref}

\usepackage{fancyhdr}
\usepackage[normalem]{ulem}
\usepackage{url}
\usepackage{tikz}
\usepackage{contour}
\usepackage{enumitem}
\usepackage{subcaption}
\usepackage{flushend}

\usepackage[ruled,linesnumbered]{algorithm2e}
\usepackage{cleveref}
\usepackage[bottom=2.1cm, top=1.20cm,left=1.72cm,right=1.72cm]{geometry}
\newcommand\Archname{The Guardian Council}
\newcommand\archname{the Guardian Council}
\newcommand\marchname{FireGuard}
\newcommand\Marchname{FireGuard}

\graphicspath{{./graphics/}}

\newcommand{\parlabel}[1]{{\noindent\bf #1}}
\newcommand{\eg}{e.g.\xspace}
\newcommand{\ie}{i.e.\xspace}

\newrobustcmd*\circled[1]{\tikz[baseline=(char.base)]{
            \node[shape=circle,draw,inner sep=0.65pt,fill,text=white,minimum size=0.95em] (char) {\textsf{\small #1}};}}

\crefname{figure}{figure}{figures}

\begin{document}

% \title{FireGuard: Comprehensive Monitoring for Superscalar Cores} % How about the tittle? Yeah fine by me - Sam
\title{FireGuard: A Generalized Microarchitecture for Fine-Grained Security Analysis on OoO Superscalar Cores}

\author{\IEEEauthorblockN{Zhe Jiang\IEEEauthorrefmark{2} Sam Ainsworth\IEEEauthorrefmark{6}, Timothy Jones\IEEEauthorrefmark{5}}
\IEEEauthorblockA{\IEEEauthorrefmark{2}National Center of Technology Innovation for EDA, School of Integrated Circuits, South East University, People's Republic of China,\\
\IEEEauthorrefmark{5}University of Cambridge, United Kingdom,
\IEEEauthorrefmark{6}University of Edinburgh, United Kingdom}
}

% \title{A Heterogeneous and Parallel Microarchitecture for Security Analysis}
% \sam{We can't have heterogeneous parallel in the title anymore, as it doesn't make sense with the current framing (where heterogeneity and parallelism are pushed to the wayside in favour of the frontend). The current one still needs work but it's along the lines we should be going for.}

\maketitle

\thispagestyle{empty}

\begin{abstract}
High-performance security guarantees rely on hardware support. 
Generic programmable support for fine-grained instruction analysis has gained broad interest in the literature as a fundamental building block for the security of future processors.
Yet, implementation in real out-of-order (OoO) superscalar processors presents tough challenges that cannot be explored in highly abstract simulators.
We detail the challenges of implementing complex programmable pathways without critical paths or contention. We then introduce \marchname, the first implementation of fine-grained instruction analysis on a real OoO superscalar processor.
We establish an end-to-end system, including microarchitecture, SoC, ISA and programming model. 
Experiments show that our solution simultaneously ensures both security and performance of the system, with parallel scalability.
We examine the feasibility of building \marchname\ into modern SoCs: Apple's M1-Pro, Huawei's Kirin-960, and Intel's i7-12700F, where less than 1\% silicon area is introduced.
The Repo. of \marchname{}'s source code:
\textbf{\url{https://github.com/SEU-ACAL/reproduce-FireGuard-DAC-25}}.
\end{abstract}
% We comprehensively examine \marchname{}'s performance, latency, overhead, scalability, and bottlenecks.
% For instance, using four microcontroller-sized checkers are sufficient to deploy a custom performance counter at only 2.5\% performance overhead, a shadow stack at 2.1\% overhead, or an AddressSanitizer-like scheme at 39\% overhead, dropping to 6\% with 12 checkers.
% Integrating hardware accelerators into \marchname 's filtering-mapping mechanisms removes performance overhead almost completely.

\section{Introduction}
\label{section:Introduction}

With ever-growing computation capacity, modern systems increasingly execute applications on shared platforms~\cite{burns2017survey,woodruff2014cheri}.
The latest in-vehicle information systems from BYD, the world's largest EV manufacturer~\cite{zecar}, allow installation and execution of less-verified third-party workloads with life-critical workloads, threatening system-wide safety and trustworthiness~\cite{checkoway2011comprehensive,wehbe2018hardware}. 
Similarly, Android phones allow untrustworthy applications to coexist with security-critical banking software~\cite{felt2011android}.
Modern systems thus need the ability to analyze, detect and mitigate vulnerabilities in an always-on and comprehensive way.

\parlabel{Existing work.}
Always-on, comprehensive security analysis relies on hardware support~\cite{bannister2019memory,vera2020inside,khasawneh2015ensemble,milenkovic2005hardware,nyman2017hardscope,kuzhiyelil2020towards,devietti2008hardbound,zeinolabedin2020real,zhang2018hcic,ozsoy2015malware}.
Current implementations, \eg, Arm's MTE~\cite{bannister2019memory} and BTI~\cite{BTI}, and Intel's LAM~\cite{LAM} and CET~\cite{CET}, have very limited flexibility, allowing an attacker to bypass them by simply shifting their targets.
Also, deployment against the latest threats requires lengthy development, leading to long vulnerability windows.

Hardware-assisted fine-grained instruction analysis~\cite{christoulakis2016hcfi,ainsworth2020guardian,fytraki2014fade,deng2010flexible,chen2008flexible} presents a new paradigm, adding observation channels into cores to filter and analyze execution (\eg, committed instructions, memory accesses and function calls), in programmable analysis engines, (\eg, microcontrollers, accelerators or FPGAs).
Through reconfigurability and parallelism, they can adapt to cover a wide range of attacks.

\parlabel{Challenges.}
{
Existing efforts have been conducted through software simulation~\cite{ainsworth2020guardian,fytraki2014fade,chen2008flexible} or on simple in-order processors~\cite{deng2010flexible,ge2017griffin}.
They do not show how to build analysis into real OoO superscalar cores, or  consider whether this is possible without a full overhaul of core design or at palatable overhead. 
When we tried to build such a mechanism into a real core, we hit bottlenecks and contention at every stage of the processing pipeline, from data collection and filtering to distribution and analysis. Typical programming models~\cite{ainsworth2020guardian} require full generality: any and all instructions can be monitored simultaneously, with any and all data selected from them, and sent to multiple analyses at once. Inside the core, such data must be collected in a narrow time window, at commit-time rather than execute-time to reflect ordering, but before the data is overwritten.
%Contentions of the data collection beyond this window necessitate a halt in the main core's execution. \sam{<- Aren't you laboring this point a bit? Surely you could just delay the freeing of the physical register?}
While data can be filtered down if it is irrelevant for currently running analyses, if every instruction could be monitored, the mechanism that decides which instructions are processed must be highly parallel and high throughput to avoid back-pressure slowing down the core.
Even when data volume has been reduced using filtering, the generality required of distribution made it a challenge for us to send data to multiple analysis engines at once without unscaleable broadcast. 
}

\parlabel{Contributions.}
We introduce the \textit{first} \textit{microarchitecture} for hardware-assisted fine-grained instruction analysis, \marchname{}, implemented in RISC-V BOOM~\cite{zhao2020sonicboom}. 
To this end, we present:

\begin{itemize}[leftmargin=*]
    \item a buffer-free data-forwarding channel by inserting bypass circuits at key locations within the main core, giving fine-grained visibility of execution without significant microarchitectural invasion;

    \item a superscalar event filter, utilizing a fleet of SRAM-based \emph{mini-filters} to handle commit of arbitrary instruction types at the same width as the core, shrinking the content volume for later analysis and preventing extra performance degradation.

    \item broadcast-free communication channels, partitioning a task mapper into a distributed fabric network and a scalable allocator. The former enables independent data paths per transaction, while the latter uses multiple \textit{Scheduling Engines} to simultaneously route data to all interested engines;

    \item a microarchitecture-assisted programming model for analysis engines, with optimizations across the hardware-software stack via new queue-communication instructions, a novel tightly coupled ISA extension (ISAX) interface to minimize data hazards, and unrolling-aware custom instructions. 
\end{itemize}

We implement FireGuard in Chisel, deploying on Virtex UltraScale+ FPGAs using FireSim~\cite{karandikar2018firesim}. 
We boot full Linux and deploy different workloads with various security schemes: a Custom Performance Counter (PMC)~\cite{ainsworth2020guardian}, a shadow stack~\cite{abadi2009control},  AddressSanitizer~\cite{serebryany2012addresssanitizer}, and Use-after-Free (UaF) detection~\cite{erdHos2022minesweeper}.
The results show that employing four analysis engines is sufficient for running a PMC with 2.5\% performance overhead, a shadow stack with 2.1\% overhead, AddressSanitizer with 39\% overhead (reducing to 6\% with 12 engines), and UaF detection with 42\% overhead (16\% with 12 engines). 
Overhead can be completely removed by further integrating hardware accelerators.
We evaluate the feasibility of building \marchname\ into modern SoCs: M1-Pro, Kirin-960 and i7-12700F, where less than 1\% silicon area is required.
% We publicly release \marchname's source code, including both hardware and software, at \textbf{\url{https://anonymous.4open.science/r/FireGuard-DAC25}}.

\begin{figure*}[t]   
    \centering
    \includegraphics[width=1\textwidth]{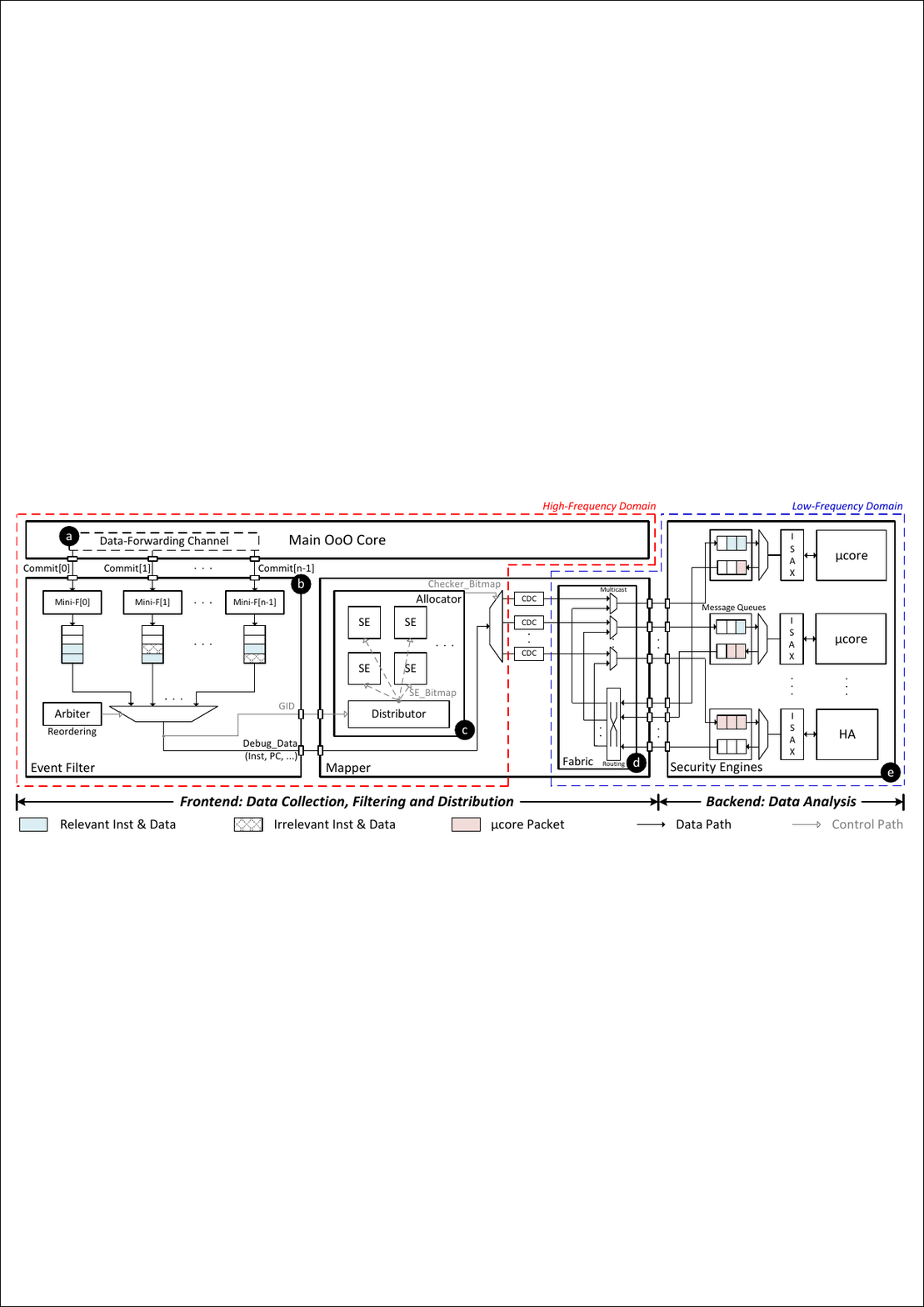}
    \vspace{-15pt}
    \caption{Overview of \marchname{} \textit{(Mini-F: Mini-Filter; GID: Group Index\protect\footnotemark; HA: Hardware Accelerator; SE: Scheduling Engine)}: 
    \circled{a} {buffer-free} data-forwarding channel extracts the main core's execution events;
    \circled{b} a {superscalar} filter pre-checks extracted events, identifying relevant instructions and selecting channels for analysis;
    \circled{c} an allocator associates an SE to each kernel to distribute contents, and \circled{d} a {distributed fabric network} transmits contents to the $\mathbf{\upmu}$cores or HAs;
    \circled{e} kernels running on $\mu$cores or HAs fetch contents and validate their security. }
    \vspace{-15pt}
    \label{fig:Top_Micro}
\end{figure*}

\vspace{-2.5pt}
\section{Fine-Grained Instruction Analysis}
\label{sc:GuardianCouncil}
\vspace{-2.5pt}
Although studies on fine-grained instruction analysis converge on the same top-level idea, the architecture and APIs across them slightly differ. 
We adopt \archname{}~\cite{ainsworth2020guardian} as a generalized example, build \marchname\ upon it and discuss the feasibility of the others.

\parlabel{\Archname{}.}
The architecture features a generic frontend that can analyze any instruction type.
It inserts a data-forwarding channel into the compute core's commit stage, collecting data (\eg,~opcode, operands) from program execution.
An event filter performs a pre-check on all observed data, selecting relevant information (dependent on the analyses being run) and sending it to the analysis engines through a mapper. 
The mapper chooses the target according to which analyses (\emph{guardian kernels}) are running on each analysis engine.
Using the passed information, the engines run guardian kernels in parallel, validating security and finding vulnerabilities.

\parlabel{The other architectures.} The key difference between \archname\ and similar mechanisms~\cite{fytraki2014fade,deng2010flexible,chen2008flexible} is the analysis engines in the backend.
It deploys a sea of microcontroller-sized cores ($\upmu$cores) to run user-programmable validations (running on conventional \textit{main cores}) in parallel. 
This allows for highly efficient, updateable and upgradeable analysis, by exploiting parallelism within validation tasks and between multiple independent tasks at once, and makes use of the fact that $\upmu$cores consume significantly less hardware, by multiple orders of magnitude, compared to OoO superscalar main cores, which must achieve high single-threaded performance unnecessary for a $\upmu$core. Deploying to an FPGA~\cite{deng2010flexible} or to multiple large analysis engines~\cite{fytraki2014fade,chen2008flexible} involves similar challenges of distribution.

%\footnotetext{Upon the guardian kernels' input demands, relevant instructions are {categorized} into different groups with a unique GID, where `0' is reserved for an irrelevant instruction.}

\section{\Marchname}
\label{section:Overview}

We show the concept of fine-grained instruction analysis is feasible by building a real system (\marchname) upon the \archname\ architecture.
% We show how to build FireGuard, a real system based on the concepts in the Guardian Council. 
\Cref{fig:Top_Micro} fleshes out the data-forwarding channel, filter, mapper and analysis engines, with a careful redesign allowing practical implementation that can handle generality at each stage\footnote{We partition the microarchitecture into two clock domains.
The main core, along with its associated modules (\eg,~L1-cache), the data-forwarding channel, the filter, and the allocator, are within a high-frequency domain driven by a fast clock source.
This avoids any resulting slowdown caused by data-forwarding, filtering, and allocating activities.
The more parallel $\upmu$cores and fabric network are within a low-frequency domain driven by a slower clock source, with handshake-based clock-domain crossing.
This ensures energy efficiency and prevents the simple $\upmu$cores from becoming the critical path.}. %-- with different data-collection and filtering strategies for each committed instruction and selective data distribution to any/all analysis engines using configurable scheduling protocols.}

\begin{figure}[!t]
    \centering
    \includegraphics[width=1\columnwidth]{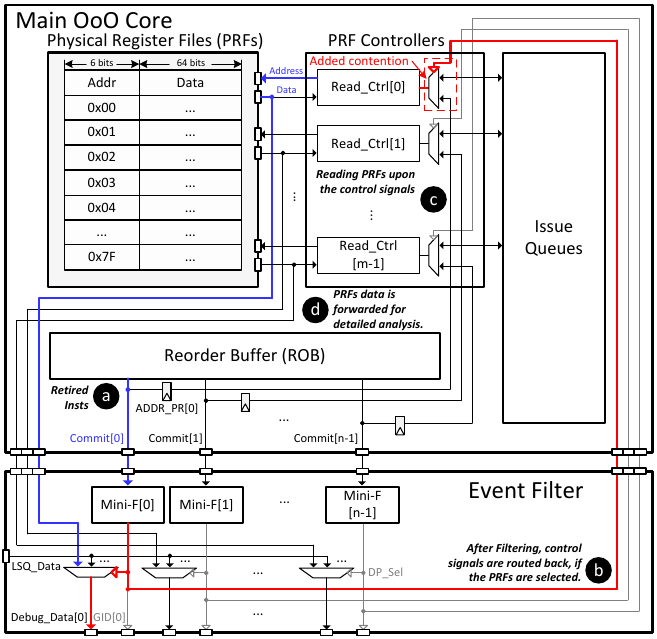}

        \caption{Data-forwarding channel, using PRFs as an example \textit{(blue lines: data-forwarding paths; red lines: filtering paths; gray lines: control paths; DP\_Sel: Data Path Selection)}:
    \circled{a} commit paths from the ROB are hooked, forwarding retired instructions to mini-filters and storing the PRF access addresses in temporary registers;
    \circled{b} mini-filters pre-check the forwarded instructions, sending control signals to PRF controllers when PRF data is selected;
    \circled{c} the data-forwarding channel preempts the controllers and feeds the addresses temporarily stored;
    \circled{d} the read data is routed back for in-depth analysis.
    }

        \label{fig:Data_ForwardingChannel}
\end{figure}

To do so, we integrate simple read-only {bypass circuits} at various locations within the main core (\cref{fig:Top_Micro}~\circled{a}) to extract data while avoiding new buffers between execute and commit.
These forward on debug data associated with committed instructions for detailed analysis.
Since the design only involves minor microarchitectural changes to the main core via adding read-only interfaces, it keeps invasion into the main core low and avoids adding significant hardware overhead.
With that, we deploy a \textit{superscalar} set of SRAM-based mini-filters to handle high commit widths in parallel (\cref{fig:Top_Micro}~\circled{b}).
It can be programmed to be sensitive to any arbitrary group of instructions, and guarantees that filtering can be completed while the data is still available in the core.
To timely deliver the data to the analysis engines, we divide the mapper into a {scalable} allocator (\cref{fig:Top_Micro}~\circled{c}) and a {distributed} fabric network (\cref{fig:Top_Micro}~\circled{d}).
The allocator associates a Scheduling Engine (SE) to each guardian kernel to independently transmit contents while avoiding broadcast, while the fabric network establishes dedicated paths for filtered contents and inter-checker communication to mitigate contention.
In each analysis engine, we develop a data-hazard-aware ISAX interface  (\cref{fig:Top_Micro}~\circled{e}) and integrate it into the Memory Access (MA) stage of the $\upmu$core's pipeline, connecting its message queues. 
This allows us to develop drivers and programming models for guardian kernels, including instructions designed for efficient, hazard-minimizing design patterns.

\vspace{-2.5pt}
\subsection{Data-forwarding channel}
\label{sc:Data-forwarding channel}
The data-forwarding channel is deployed at the main core's commit stage, extracting, selecting, and transporting debug data associated with every retired instruction.  
Since the relevant data is already stored in different locations within the main core, we design a buffer-free implementation (\ie,~adding no new intermediate storage between out-of-order execute and in-order commit), by inserting bypass circuits at the Reorder Buffer (ROB), Physical Register Files (PRFs), Load Store Queue (LSQ), and Fetch Target Queue (FTQ). This transports PC address, instruction data, operand data, and memory and jump addresses during commit for any instruction selected by the filter, avoiding reads of information not selected.
These points cover all locations that store outputs of arithmetic-logic units, enabling fine-grained visibility with minimal contention.

\parlabel{PRF Example.}
\Cref{fig:Data_ForwardingChannel} shows the microarchitecture using PRFs as an example, connecting the ROB, filter, and PRFs. On the ROB side, we add logic to hook onto each commit, transmitting the retired instructions to the mini-filters, as well as address registers storing the PRF indices accessed by each instruction (\cref{fig:Data_ForwardingChannel}~\circled{a}).
In the cycle following retirement, the mini-filters identify Group Indexes (GIDs) of the transmitted instructions and select data based on programmed settings (section~\ref{sc:Event Filter}). 
If the PRF data is selected, a control signal is routed back, preempting the PRF controller (\cref{fig:Data_ForwardingChannel}~\circled{b}).

The PRF's read controllers are statically multiplexed (\cref{fig:Data_ForwardingChannel}~\circled{c}) between the issue queue (for executing instructions) and the mini-filters. Mini-Filter\verb|[x]| has priority access to Read\_Ctrl\verb|[x]| should it require it, to allow the data to be read and transported immediately (\cref{fig:Data_ForwardingChannel}~\circled{d}), avoiding any buffering or delays in freeing the physical register. This means that an instruction attempting to use the same port will be delayed until the next cycle, resulting in contention\footnote{
In LDQ, STQ, and FTQ, similar microarchitectures are deployed to obtain memory access or jump addresses. 
Unlike PRFs, where forwarded data can be stored at arbitrary addresses, the tops of these queues consistently hold the data associated with the most recently retired instructions. 
When a mini-filter decides to forward a load, store, or jump address, the bypass circuits directly transmit from the relevant queue's top, avoiding contention.}.

\begin{figure}[t]
    \centering
    \includegraphics[width=1\columnwidth]{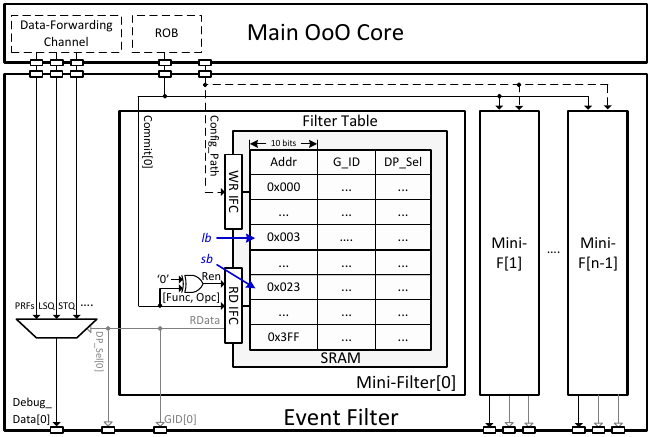}
    \vspace{-15pt}
    \caption{Microarchitecture of a mini-filter \textit{(gray lines: control paths; WR/RD IFC: Write/Read Interface)}.
    }
    \vspace{-15pt}
    \label{fig:Mini_Filter}
\end{figure}

\subsection{Event Filter}
\label{sc:Event Filter}
The event filter pre-checks all retired instructions.
If instructions are selected for analysis, it returns their Group Indexes (GIDs) for the mapper and programs the data-forwarding channel to select data. 
We give a superscalar implementation that enables simultaneous filtering of all instructions,
ensuring the filter keeps up with the main core.

\label{sbsc:filter microarchitecture}
\Cref{fig:Top_Micro} \circled{b} shows its design, including a set of mini-filters, FIFO queues, and an arbiter.
A mini-filter is connected to each superscalar commit path of the ROB. Each is indexed by the instruction opcode from the data-forwarding channel, and returns programmed GIDs and selects debug data from the chosen channel for instructions analyzed.
Filtered contents are buffered into paired FIFO queues, allowing a shared arbiter to arrange the output into sequence.

\begin{figure}[!h]
    \centering
    \vspace{-8pt}
    \includegraphics[width=1\columnwidth]{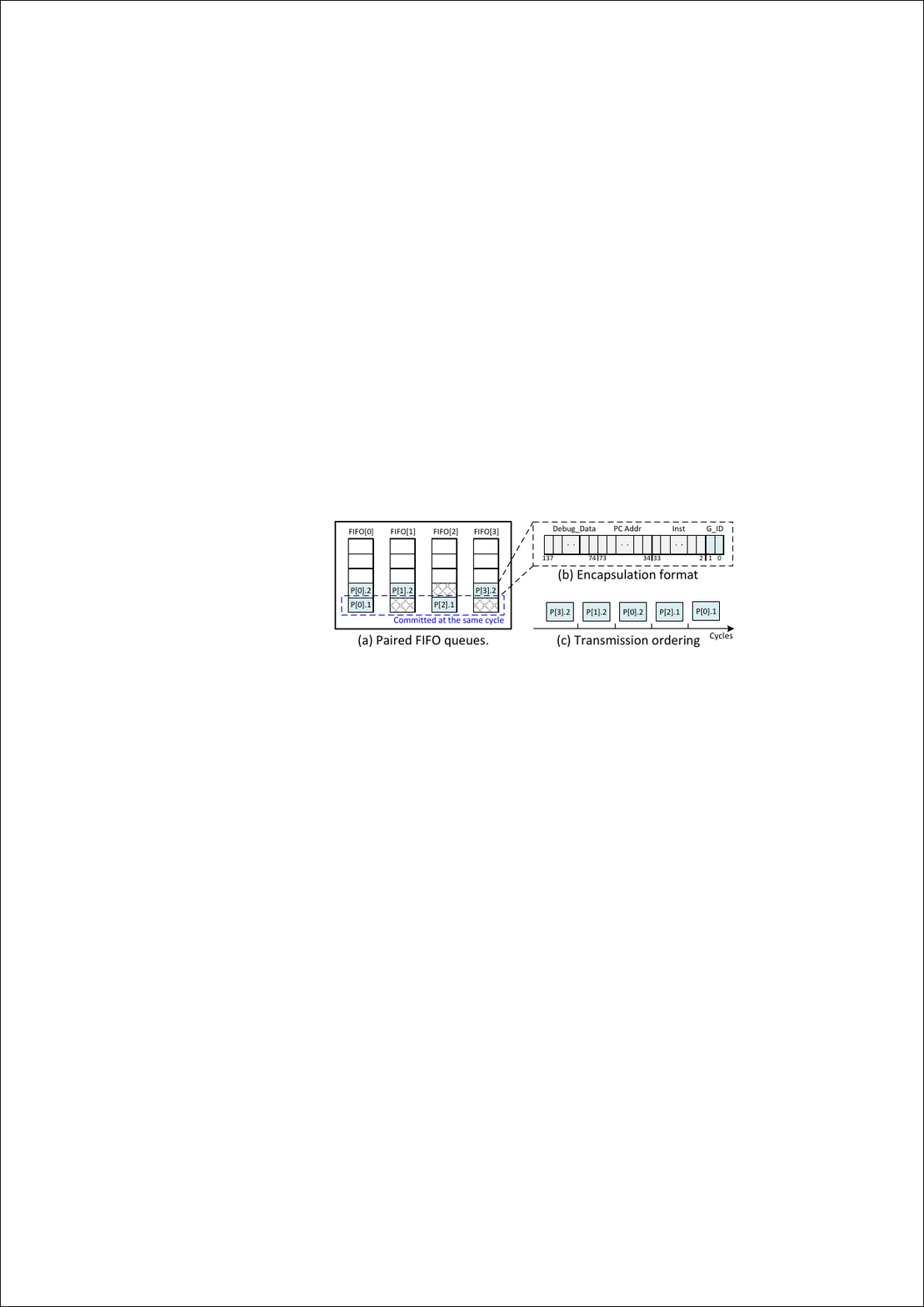}
    \vspace{-14pt}
    \caption{Reordering with 4-width FIFOs. P[x].y: yth packet in FIFO[x].}
    \vspace{-4pt}
    \label{fig:Re-ordering}
\end{figure}

\parlabel{Mini-filters.}
Mini-filters use a SRAM-based look-up table (\cref{fig:Mini_Filter}). 
The address (10-bit) is an index formed of the concatenated RISC-V opcode (lower 7 bits) and function code (higher 3 bits) for all possible instructions, and stores the mapper's GID and the desired data paths (PRF, LSQ and/or FTQ) for each instruction. %\footnote{Multiple data paths can be chosen for one instruction if it is collectively required, consuming multiple cycles due to the transmission format (\cref{fig:Re-ordering}(b)).}.
For instance, 0x03 and 0x23 index RISC-V \texttt{lb} and \texttt{sb}, respectively.
We route the instructions' opcodes and function codes to the address port of the SRAM's read interface, and direct the read data to the data-forwarding channels and the mapper, assisting in data selection and allocation\footnote{We place a pair of FIFO queues to connect the mini-filters, buffering the filtered contents (\cref{fig:Re-ordering}(a)).
Although these filtered contents are produced by the mini-filters in parallel, they must be sent sequentially, aligning with their commit order, since analyses can be sensitive to program order (\eg, shadow stack~\cite{abadi2009control}).
We encapsulate filtered contents to achieve this (\cref{fig:Re-ordering}(b)). 
%If an instruction is identified as relevant, the results are encapsulated and pushed into the FIFO queue. 
If an instruction is discarded, an invalid packet is generated and also pushed into the FIFO queue in order to preserve ordering at the end of the FIFO.
The arbiter uses a finite state machine to transmit packets (\cref{fig:Re-ordering}(c)) in-order, consuming one clock cycle for a valid packet while skipping invalid packets.}.

\vspace{-5pt}
\subsection{Mapper}
\label{sc:Mapper}
\vspace{-2pt}

The mapper routes filtered packets to engines based on configured parallelization policies, and also enables data exchange between the analysis engines, fostering complex parallelism schemes for guardian kernels.  It is fully programmable, allowing any instruction to flow to all interested guardian kernels and be scheduled to an analysis engine core for each. \Cref{fig:Top_Micro} \circled{c} and \circled{d} illustrate the microarchitecture of the mapper, which we partition into a scalable \textit{allocator} followed by a distributed \textit{fabric network}.
The mapper transitions \marchname's processing from superscalar to scalar: unlike the filter, it only handles one packet per cycle. 
This rarely impedes a 4-wide BOOM's performance (we saw less than 0.5\% slowdown)\footnote{Even so, a superscalar mapper could be considered for a more powerful core.
To achieve this, modifications are necessary for both the fabric and allocator, including duplicating communication channels and SEs.
Extra arbiters must be deployed to manage contention, \eg, when multiple packets are sent to the same security engine.}.

\begin{figure}[t]
    \centering    \includegraphics[width=1\columnwidth]{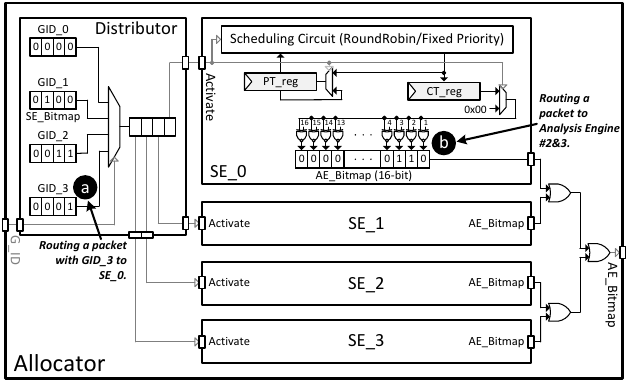}
    \vspace{-15pt}
    \caption{Allocator microarchitecture. 3 GIDs and 4 SEs allocate contents to 16 engines  \textit{(gray lines: control paths)}.}
    \vspace{-15pt}
    \label{fig:Allocator}
\end{figure}

\begin{figure*}[t]
    \centering
    %No vspace before figures -- goes out of margins
    \includegraphics[width=1\linewidth]{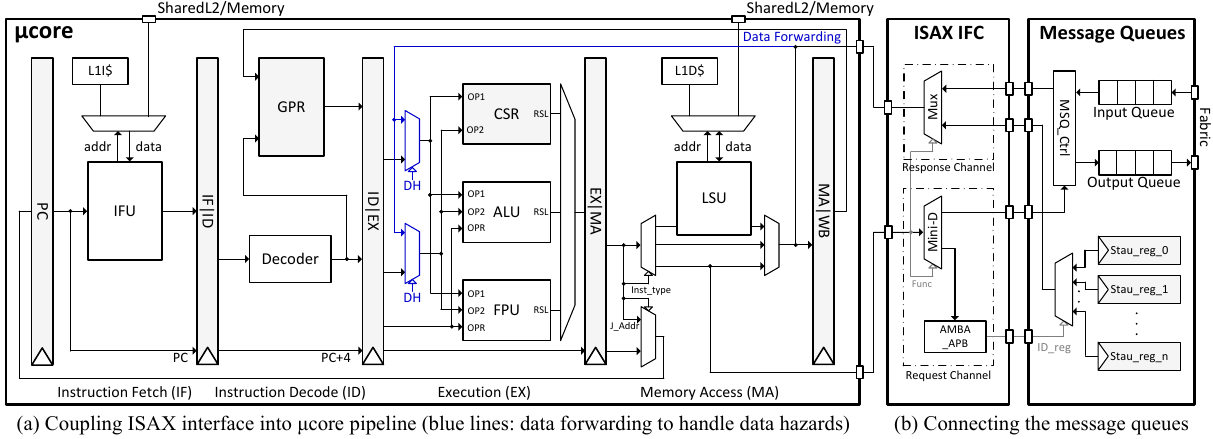}
    \vspace{-15pt}
    \caption{Custom instructions are integrated into the $\mu$core's pipeline at the MA stage. \textit{(gray lines: control paths; MSQ\_Ctrl: Message Queue Controller;
    IFU: Instruction Fetch Unit; Mini-D: Mini-Decoder):}
    For the ISAX interface (IFC), a full-duplex microarchitecture handles requests and responses independently, with a mini-decoder and multiplexer. 
    In the request channel, a mini-decoder directs an ISAX request to the message queues' MSQ\_Ctrl or status registers (via an APB bridge).
    In the response channel, a multiplexer selects returned data, routing it to the commit stage and the EX stage using forwarding logic for handling data hazards. }
    \vspace{-10pt}
    \label{fig:ISAX}
\end{figure*}

\parlabel{Allocator.}
The allocator uses a 2-level indirection bitmap to allocate packets across the analysis engines.
\Cref{fig:Allocator} shows the allocator, including a distributor and a set of Scheduling Engines (SEs).
The distributor manages a bitmap between GIDs and SEs, deciding which SE(s) should be activated during packet transmission.
SEs are \emph{one-to-one} associated to a guardian kernel, and each maintains \textit{another} bitmap between itself and analysis engines, allocating filtered packets to groups of analysis engines  executing one guardian kernel.

In the distributor, an SE$\_$Bitmap register is assigned to each GID, and individual register bits are used to index the SEs that are interested in that specific GID. 
For instance, if packets with GID 3 should be sent to SE 0, bit 0 in SE$\_$Bitmap[3] is set (\cref{fig:Allocator}~\circled{a}).
%We connect the output of the SE$\_$Bitmap registers to input data ports of the multiplexer and direct the GID given by the event filter to the multiplexer's control port.
SEs use a scheduling circuit, two scheduling registers (PT$\_$reg and CT$\_$reg), and an AE$\_$Bitmap (Analysis Engine Bitmap) register.
The scheduling circuit implements several policies, \eg,~fixed, round-robin, and block mode~\cite{ainsworth2020guardian}, where the latter is used to send all messages to one $\upmu$core until it is full before moving to the next, for when message locality is important, \eg,~shadow stack. %, ensuring that each scheduling decision is consistently made in a fixed clock cycle.
When the scheduling circuit is activated, the PT$\_$reg (\textit{previous target}) is used to generate the \textit{current target}, buffered in CT$\_$reg until it is moved to the PT$\_$reg after packet transmission. 
The CT$\_$reg value sets the relevant target bit in the SE's AE$\_$Bitmap (\cref{fig:Allocator}~\circled{b}).
The AE$\_$Bitmaps returned by all SEs are combined using OR gates to form the decision, selectively broadcasting to the analysis engines.
% For example, if the SEs return "0x50", the packet will be transmitted to checkers with indices 5 and 7.

%\begin{figure*}[!t]
%    \centering
%    \includegraphics[width=1\textwidth]{graphics/ISAX.pdf}
%    \vspace{-14.5pt}
%    \caption{Custom instructions are integrated into the $\upmu$core's pipeline at the MA stage. \textit{(gray lines: control paths; MSQ\_Ctrl: Message Queue Controller;
%    IFU: Instruction Fetch Unit; Mini-D: Mini-Decoder):}
%    For the ISAX interface (IFC), a full-duplex microarchitecture handles requests and responses independently, with a mini-decoder and multiplexer. 
    %In the request channel, a mini-decoder directs an ISAX request to the message queues' MSQ\_Ctrl or status registers (via an APB bridge).
    %In the response channel, a multiplexer selects returned data, routing it to the commit stage and the EX stage using forwarding logic for handling data hazards. 
%    }
%    \vspace{-12.5pt}
%    \label{fig:ISAX}
%\end{figure*}

\parlabel{Fabric network.}
The fabric network features a half-duplex multicast (1-to-N) channel and a full-duplex routing (N-to-N) channel.
The multicast channel selectively broadcasts packets, while the routing channel allows the checkers to transmit packets among themselves.

The multicast channel uses multiplexers to direct packets from the event filter to the message queues in the analysis engines. 
All multiplexers are controlled by the allocator, regulating transmission and masking of each packet.
The routing channel uses a Manhattan grid~\cite{ankur2006noc,schoeberl2015t,jiang2018bluevisor,jiang2022bluescale} Network-on-Chip (NoC) mesh.
Each router has five bi-directional ports, connected to other routers located to its north, south, east, and west, as well as to an analysis engine. 

\begin{table}[t]
\vspace{3pt}
\resizebox{.95\columnwidth}{!}{%
\begin{tabular}{c|c|l}
\bottomrule
\hline
\textbf{Instruction} & \textbf{\begin{tabular}[c]{@{}c@{}}Target\\ Queue\end{tabular}} & \textbf{Descriptions}                                                                                                    \\ \hline
\texttt{count rd, rs1}        & \begin{tabular}[c]{@{}c@{}}Input/\\ Output\end{tabular}         & \begin{tabular}[c]{@{}l@{}}Count number of packets buffered in \\ message queue rs1, returning it to rd. \end{tabular}                        \\ \hline
\texttt{top rd, rs1}          & Input                                                           & \begin{tabular}[c]{@{}l@{}}Return bitfields {[}rs1+63:rs1{]} of the\\ first element to rd.\end{tabular}                  \\ \hline
\texttt{pop rd, rs1}          & Input                                                           & \begin{tabular}[c]{@{}l@{}}Remove the first element and return \\ its bitfields {[}rs1+63:rs1{]} to rd.\end{tabular}     \\ \hline
\texttt{recent rd, rs1}       & Input                                                           & \begin{tabular}[c]{@{}l@{}}Return bitfields {[}rs1+63:rs1{]} of the \\ most recently removed element to rd.\end{tabular} \\ \hline
\texttt{push rs1}             & Output                                                          & Push packet rs1 for transmission.                                                                                        \\ \hline
\toprule
\end{tabular}
}
\vspace{-2pt}
\caption{Main control instructions for the message queues.}
\vspace{-10pt}
\label{table:ISA}
\end{table}

\subsection{ISA and Programming Model}
\label{sc:Security chckers}
The analysis engines run guardian kernels concurrently on $\mu$cores, finding vulnerabilities. 
To handle the frequent hardware-software interactions, we use a FIFO-based programming model implemented into RISC-V Rocket cores via custom instructions. 
As these instructions take up a large fraction of total $\mu$core cycles, we integrate FIFO-management instructions using a new tight-coupling arrangement; Rocket's existing ISAX interface, which executes custom instructions post-commit, caused too many data hazards. 
We also redesign our instruction set to better support high-throughput programming design patterns that minimize hazards further.

\parlabel{ISA.} 
We connect the message queues to the fabric network in the mapper, allowing the $\mu$core to receive packets via an input queue and send packets through an output queue, in order to support pipelined parallelism strategies such as used in the shadow stack~\cite{ainsworth2020guardian}.
Our queue controller, with status registers and software drivers, allows guardian kernels to manage the message queues using ISAX~\cite{asanovic2016rocket} custom instructions (see table~\ref{table:ISA}). \texttt{Top}, \texttt{push} and \texttt{pop} were inherited from the original design~\cite{ainsworth2020guardian}; \texttt{count} was added to aid in loop unrolling, and \texttt{recent} to allow accessing extra information about an element already processed: for example, the PC is only needed on a detected error in AddressSanitizer, and is discarded otherwise.

\parlabel{Microarchitecture Support.} 
RISC-V Rocket runs custom instructions post-commit~\cite{asanovic2016rocket}.
The routing to the ISAX peripheral blocks the core for at least 3 cycles 
for each instruction and can extend up to 13 cycles in the presence of data hazards and contention~\cite{Pala2017rocc}.
This causes data hazards and large slowdowns when such instructions are as commonly used as our queue operations. 
We redesigned Rocket's interface to move it into the MA stage\footnote{The MA stage was chosen as it is the first stage in the pipeline that is non-speculative, simplifying the implementation of the state-destructive \texttt{pop} but otherwise minimizing hazards, such that only one bubble is required for an instruction immediately following the custom instruction that uses its output.} of the pipeline (\cref{fig:ISAX}), multiplexing between the ISAX unit and the load-store unit.

\begin{table}[t]
\centering
\resizebox{1\columnwidth}{!}{%
\begin{tabular}{ll}
\multicolumn{2}{c}{\textit{Main core}}                                                                                \\ \hline
Core                                                        & 4-Width, out-of-order SonicBOOM~\cite{zhao2020sonicboom}, @3.2GHz                                                                                                                                                       \\
Pipeline                                                    & \begin{tabular}[c]{@{}l@{}}128-Entry ROB, 96-Entry IQ, 32-entry LDQ/STQ, \\ 128 Int/FP Phy Registers, 2 Int ALUs, 1 FP/Multi/Div \\ ALU, 2 MEM, 1 Jump Unit, 1 CSR Unit\end{tabular} \\
\begin{tabular}[c]{@{}l@{}}Branch \\ Predictor\end{tabular} & \begin{tabular}[c]{@{}l@{}}TAGE algorithm, 256-entry BTB, 32-entry RAS,\\ 6 TAGE table with 2 - 64 bits history\end{tabular}                                                         \\
\multicolumn{2}{c}{\textit{Memory}}                                                                                                                                                                                                                \\ \hline
L1 ICache                                                   & 32KB, 8-way, 8 MSHRs                                                                                                                                                                 \\
L1 DCache                                                   & 32KB, 8-way, 8 MSHRs                                                                                                                                                                 \\
L2 Cache                                                    & 512KB, 8-way, 12 MSHRs                                                                                                                                                               \\
LLC                                                         & 4MB, 8-way, 8 MSHRs                                                                                                                                                                  \\
Memory                                                      & 16 GB DDR3 @1066MHz, max 32 requests                                                                                                                                              \\
\multicolumn{2}{c}{\textit{FireGuard and Interconnects}}                                                                                                                                                                                           \\ \hline
Event Filter                                                & 4-width, 16-entry FIFO                                                                                                                                                               \\
Mapper                                                      & 4 SEs, 8-entry CDC, fabric @1.6GHz                                                                                                                                                   \\
Analysis Engine                                                     & \begin{tabular}[c]{@{}l@{}}In-order Rocket $\mu$core~\cite{asanovic2016rocket}, 5-stage pipeline, @1.6GHz, \\ 32-entry message queues, no FPU\end{tabular}                                                                              \\
L1 Cache                                                    & 4KB, 2-way for both I- and D-Cache                                                                                                                                                   \\
Interconnect                                                & Memory bus @ 1GHz, others @ 3.2GHz                                                                                                                                                   
\end{tabular}
}
\caption{Hardware configurations evaluated.}
\label{table:HardwareSetup}
\end{table}

\begin{figure*}[t]
    \centering
    \includegraphics[width=.9\linewidth]{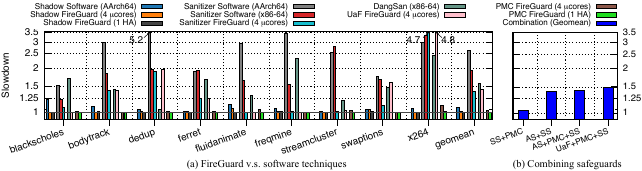}

        \caption{Performance results for \marchname\ \emph{(4 $\upmu$cores or 1 HA for each kernel)} running Parsec with different/combined safeguards
    \emph{(AS: AddressSanitizer; SS: Shadow Stack; in (b), SS is implemented as a HA when three guardian kernels are deployed).
    }}

        \label{fig:performance}
\end{figure*}

\section{Evaluation}
\label{sc:Evaluation}

\parlabel{Experimental Setup}
To comprehensively evaluate the feasibility of \marchname, we built \marchname{} into RISC-V cores. BOOM main cores were augmented with data-forwarding channels, filters, and mappers; 
Rocket $\mu$cores were configured as security engines, running different safeguards, \ie, kernels.
% Rocket $\upmu$cores were given new custom instructions for queue management, and a highly modified custom-instruction interface that triggers fewer data hazards than the default (which handles them post-commit instead of in-pipeline).  
We implemented the microarchitecture with Chisel (v3.4) and synthesized the RTL using Vivado toolchains (v2021.2). The generated netlist was deployed on Virtex UltraScale+ FPGAs using FireSim~\cite{amid2020chipyard,karandikar2018firesim}, emulating the setup in \cref{table:HardwareSetup}.

We booted Linux (with kernel v5.7.0) and executed Parsec~\cite{bienia2008parsec}, running the simmedium dataset with kernels: Custom Performance Counter with bounds check (PMC)~\cite{ainsworth2020guardian}, AddressSanitizer~\cite{serebryany2012addresssanitizer}, Use-after-Free detector (UaF)\footnote{Our design takes MineSweeper~\cite{erdHos2022minesweeper} and uses analysis engines to find loads and stores to quarantined regions, to find as well as prevent bugs.} and shadow stack~\cite{ss_llvm,abadi2009control}.

\begin{figure}[t]
    \centering
    \includegraphics[width=0.85\linewidth]{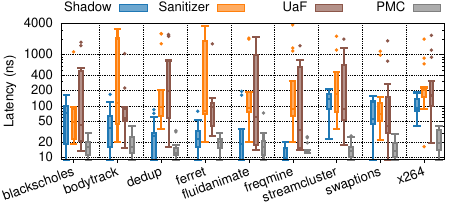}

        \caption{Detection latency while using 4 $\upmu$cores \emph{(unit: ns)}.}

        \label{fig:latency}
\end{figure}

\subsection{Performance Overhead}
\label{sbsc:PerformanceOverhead}
Figure~\ref{fig:performance}(a) shows the slowdown experienced by the main core while executing kernels in \marchname, in comparison to software-based counterparts (implemented via LLVM). 
In all instances, \marchname\ is configured with four $\upmu$cores. 
Hardware Accelerators (HAs) within \marchname\ are also presented for PMC and shadow stack.

Using 4 $\upmu$cores is sufficient to run a PMC with overheads of 2.5\% geomean,  shadow stack at 2.1\%, AddressSanitizer at 39\%, or UaF at 42\%. 
By replacing the $\upmu$cores with a single HA, the overhead of PMC and shadow stack can be reduced to 0\%.
Software techniques give 7.9\% overhead for shadow stack, and 163.5\% for AArch64 and 91.5\% for x86-64 architectures on AddressSanitizer\footnote{
AddressSanitizer is not supported for RISC-V in either GCC or LLVM, so the comparison focuses on AArch64 and x86-64. For shadow stack, support is absent for both RISC-V and x86-64, so we compare against AArch64.}. 
\marchname\ consistently outperforms software techniques, with the exception of x264 with AddressSanitizer, and Dedup on UaF. The former is due to the exceptionally high volume of load and store instructions executed in x264, where four $\upmu$cores fail to keep up to such a degree that sharing compute resources with the compute core, as in software schemes, is favorable (though \marchname\ can do better with more $\upmu$cores, see \cref{sbsc:PerformanceScalability}). The latter is caused by DangSan~\cite{dangsan} being more suited to dedup's allocation patterns than our UaF detector's underlying Minesweeper~\cite{erdHos2022minesweeper}, where DangSan itself on \marchname\ would give false negatives, as it relies on zeroing pointers, which may occur only after the UaF access if offloaded to a $\upmu$core.

\parlabel{Combining kernels.}
Figure~\ref{fig:performance}(b) reports the slowdown experienced by the main core while executing multiple kernels simultaneously.
The kernel that incurs the most slowdown dominates the performance, but slowdowns are not multiplied.
This is benefited from the parallel execution of the kernels. 
When one kernel halts the execution of the main core, decreasing the number of instructions, other kernels can persist in their function with reduced computational overhead.

\subsection{Detection Latency}
\label{sbsc:Detection Latency}
% \hugo{It is strange to inject faults in a security paper, ``pollute'' data? \eg, for shadowstack, ASAN ... \\}
% 
% To examine the detection latency, we developed an Error Injection Unit (EIU) and deployed it at the data-forwarding channel, \emph{polluting} the contents forwarded to the $\upmu$cores.

To examine the detection latency, we inject erroneous input at various locations in the core, e.g., the jump unit, the LDQ and the STQ, etc, simulating e.g., jump to a hijacked PC and access a freed memory address.
For each workload, 50--100 attacks are generated, and the detection latencies are shown in \cref{fig:latency}. 
PMC experiences the lowest detection latency, consistently detecting attacks under 50 $ns$. 
Due to its complex block parallelism model~\cite{ainsworth2020guardian}, the shadow stack sees marginally higher latencies than PMC.
In the worst-case scenario (x264), a latency of 220ns is observed.
For AddressSanitizer, though the median is always $<$200ns, the tail exceeds 2,000ns. This occurs when TLB and cache misses co-occur on many accesses in the same queue: our accurate modeling of TLB misses in FireSim causes worst-case delays to be higher than in the original work~\cite{ainsworth2020guardian}. 
% As with the Guardian Council~\cite{ainsworth2020guardian}, we mitigate security concerns by ensuring that the guardian kernels cannot be terminated or reconfigured, and system calls not be executed, until no in-flight checks remain.

\begin{figure}[t]
    \centering
    \includegraphics[width=.85\linewidth]{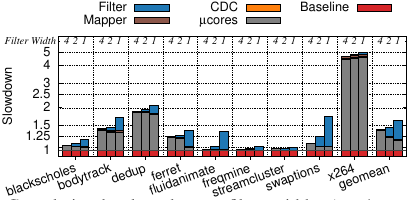}
    \caption{Cumulative bottlenecks v.s. filter widths (on 4-$\upmu$core Sanitizer), measured by the proportion of time queues are full.
    }
    \label{fig:backpressure}
\end{figure}

\begin{figure*}[t]
    \centering
    \includegraphics[width=.95\linewidth]{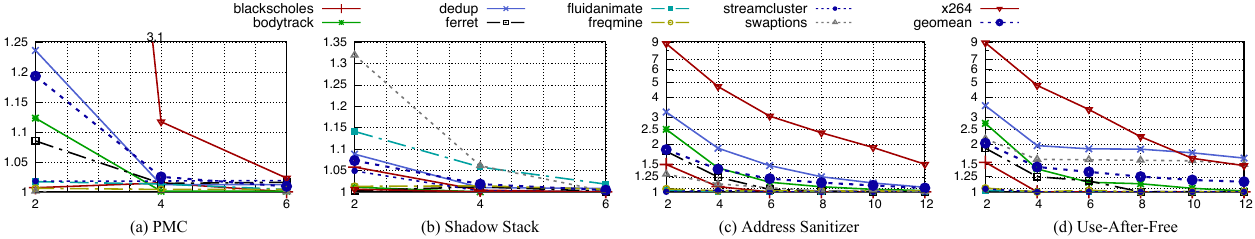}
    \caption{Slowdown when using varying numbers of $\mu$cores for guardian kernels \emph{($x$-axis: number of $\mu$cores; $y$-axis: slowdown)}.}
    \label{fig:scalability}
\end{figure*}

\begin{figure}[t]
    \centering
    \includegraphics[width=.85\linewidth]{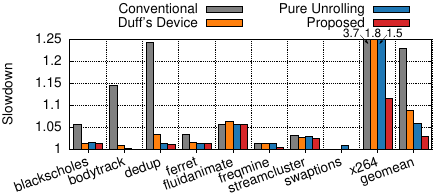}
    \caption{Slowdown of programming models (4-$\mu$core PMC).
    }
    \label{fig:programmingmodel}
\end{figure}

\subsection{Analysis of Microarchitecture Bottlenecks}
\label{sbsc:Analysis on Design Trade-offs}
Figure~\ref{fig:backpressure} decomposes overheads in a 4-$\upmu$core \marchname\ when running AddressSanitizer using varying event-filter widths (1, 2, or 4), highlighting the accumulated bottlenecks stemming from different elements.
A 4-width event filter is as wide as the core's commit, so can always keep up. 
As the width decreases, overhead increases significantly, from 16\% at 2-width to 34\% at 1-width.
The geomean overhead from the filter jumps from 16\% (2-width) to 34\% (1-width).

% By contrast, the remaining bottlenecks, from the 1-width mapper and clock-domain crossing are negligible: compute overhead of the $\upmu$cores dominates given sufficient filter parallelism. Likewise, the in-core overhead from the data-forwarding channel sharing PRF ports with the core only adds 0.3\% overhead on shadow stack (not shown, since Sanitizer accesses the LDQ instead and sees no loss).

%\{-3pt}
%\subsection{Analysis of Programming Models}
%\label{sbsc:Analysis-of-ProgrammingModels}
%\{-4pt}

%Figure~\ref{fig:programmingmodel} shows the performance overhead when executing PMC using different programming models with 4 $\upmu$cores, see section~\ref{sbsc:ProposedPM}. 
%By being aware of the remaining hazards in the $\upmu$cores, we can improve performance significantly. 
%A conventional single-iteration loop suffers from frequent data hazards due to both queue \texttt{nonempty} checks and \texttt{pop} instructions. Duff's device reduces hazards on size checks, made possible by the queue \texttt{count} instruction, and pure unrolling helps further, by removing hazards for \texttt{pop} instructions if the queue is relatively full. A hybrid strategy, using unrolling when possible and Duff's device otherwise, is uniformly best.
%\tim{I see that `count' is in code font in this paragraph.  Should we do that for `pop' and all other instruction names when they are used throughout?}\hugo{Sorry Tim, I don't get your points.}

\subsection{Analysis of Scalability}
\label{sbsc:PerformanceScalability}
Figure~\ref{fig:scalability} shows the slowdown with varying $\upmu$core count.

\parlabel{PMC.} 
Figure~\ref{fig:scalability}(a) shows that using two $\upmu$cores to perform PMC leads to a 20\% geomean slowdown. %, primarily due to a high number of load operations. 
Four $\upmu$cores reduce this to $2\%$; yet, x264 still has a 17\% slowdown. 
When scaling the  $\upmu$cores number to six, all workloads can be run with an overhead of less than 5\%.

\parlabel{Shadow stack.}
Figure~\ref{fig:scalability}(b) shows that using two $\upmu$cores for executing shadow stack leads to 7.3\% slowdown (geomean), which is comparable to the performance of LLVM's software mechanism. 
With four $\upmu$cores, the geomean overhead is reduced to 2.1\%, decreasing to 0.4\% when using six $\upmu$cores. 
With six $\upmu$cores, all workloads can be executed with $<$1\% slowdown, except fluidanimate
(1.8\%).

\parlabel{AddressSanitizer.}
Figure~\ref{fig:scalability}(c) is heavier than (a) or (b). Using two $\upmu$cores results in 86\% overhead, still outperforming software (section~\ref{sbsc:PerformanceOverhead}). 
However, three workloads (bodytrack, dedup, x264) exhibit over 100\% overhead. This is due to the extremely high volume of load and store instructions that necessitate analysis.
%When increasing the number of $\upmu$cores to six, most workloads experience less than a 5\% slowdown. However, the geometric mean overhead remains at 14.4\%, influenced by the same three workloads.
Even with 12 $\upmu$cores, x264 continues to endure a 58.9\% slowdown, while all other workloads see negligible overhead (less than 1\%).

\parlabel{UaF.} Figure~\ref{fig:scalability}(d) is the heaviest workload. As well as monitoring all loads and stores, the extra work to quarantine and release memory allocations only when safe~\cite{erdHos2022minesweeper} causes some overheads that do not parallelize away; for example, dedup sees flat overheads due to extra memory-allocation work. 
The geomean reaches 1.16$\times$ by 12 $\upmu$cores, lower than the 1.6$\times$ that is achieved by software mechanisms~\cite{dangsan}.

% \subsection{Analysis of Programming Models}
% \label{sc:APM}
% Figure~\ref{fig:programmingmodel} shows the performance overhead when executing PMC using different programming models (see~\cref{sc:Security chckers}) with 4 $\mu$cores. 
% By being aware of the remaining hazards in the $\mu$cores, we can improve performance significantly. 
% A conventional single-iteration loop suffers from frequent data hazards due to both queue \texttt{count} checks and \texttt{pop} instructions. 
% Duff's device reduces hazards on size checks, made possible by the queue \texttt{count} instruction, and pure unrolling helps further, by removing hazards for \texttt{pop} instructions if the queue is relatively full.
% A hybrid strategy, using unrolling when possible and Duff's device otherwise, is uniformly best.

\subsection{Analysis of Programming Models}
\label{sc:APM}
Figure~\ref{fig:programmingmodel} shows the performance overhead when executing PMC using different programming models (\cref{sc:Security chckers}) with 4 $\mu$cores. 
By being aware of the remaining hazards in the $\mu$cores, we can improve performance significantly. 
A conventional single-iteration loop suffers from frequent data hazards due to both queue \texttt{count} checks and \texttt{pop} instructions. 
Duff's device~\cite{duff1988duff} reduces hazards on size checks, made possible by the queue \texttt{count} instruction, and pure unrolling helps further, by removing hazards for \texttt{pop} instructions if the queue is relatively full.
A hybrid strategy, using unrolling when possible and Duff's device otherwise, is uniformly best.

\subsection{Hardware Overhead}
\label{sbsc:HardwareOverhead}
We study hardware overhead of the microarchitecture by performing a physical implementation of a 4-$\upmu$core \marchname\ at the post-layout stage using Synopsys 14$nm$ Generic PDKs. 
The RTL is synthesized with Design Compiler (v2022.12), and the netlist is placed and routed via IC Compiler 2 (v2022.12). 
% The target frequencies for the high- and low-frequency domains are set to 2 GHz and 1 GHz, respectively.

% \begin{figure}[t]
%    \centering
%    \{-5pt}
%    \includegraphics[width=.98\linewidth]{plots/bottlenecks.eps}
%  \{-15pt}
%    \caption{Cumulative bottlenecks v.s. filter widths (on 4-$\upmu$core Sanitizer), measured by the proportion of time queues are full.
%    }
%    \{-12pt}
%    \label{fig:backpressure}
%\end{figure}

The area of the SoC is 2.91$mm^2$, where the BOOM is 1.107$mm^2$ and each Rocket is 0.061$mm^2$.
The total area of \marchname 's transport mechanisms is 0.043$mm^2$ (3.88\% and 1.48\% of the BOOM and the SoC): the filter occupies 0.032$mm^2$ and the mapper 0.011$mm^2$.
Building a 4-$\upmu$core \marchname\ upon a BOOM thus needs 0.287$mm^2$, i.e.,~25.9\% and 9.86\% of the BOOM and the SoC.

Intuitively, this observation diverges from the goal of low-overhead design.
But, this is only because the prototype BOOM cores are small compared with commercial designs, yet the Rocket cores are very large, with closed-source in-order 64-bit cores available using half the resources~\cite{BTI}.
To provide more comprehensive insights, we present a detailed analysis below on real-world SoCs.
% Taking these points into account, it remains promising to build a real system using \archname\ architecture with acceptable hardware overheads.

%\begin{figure}[t]
%    \centering
%    \includegraphics[width=.9\linewidth]{plots/programmingmodel.eps}
%        \{-11pt}
%    \caption{Slowdown of programming models (4-$\upmu$core PMC).
%    }
%    \{-15.5pt}
%    \label{fig:programmingmodel}
%\end{figure}

% \begin{figure}[t]
%     \centering
%    \includegraphics[width=1\linewidth]{plots/programmingmodel.eps}
%    \{-17pt}
%    \caption{Slowdown of programming models (4-$\mu$core PMC).
%    }
%    \{-15pt}
%    \label{fig:programmingmodel}
%\end{figure}

\begin{table}[t]
\resizebox{1\columnwidth}{!}{%
\centering
\small
\begin{tabular}{lllll}
\bottomrule
\hline
\multicolumn{5}{c}{\textit{Processors (Performance Core only)}}                                                                                                                                                                                                                                                                                                                                          \\ \hline
\multicolumn{1}{l|}{\textbf{\begin{tabular}[c]{@{}l@{}}Core\\ (SoC)\end{tabular}}}            & \textbf{BOOM\cite{zhao2020sonicboom}}                                                          & \textbf{\begin{tabular}[c]{@{}l@{}}FireStorm\cite{m1pro-micro}\\ (M1-Pro)\end{tabular}}  & \textbf{\begin{tabular}[c]{@{}l@{}}Cortex-A76\cite{a76-arch}\\ (Kirin-960)\end{tabular}} & \textbf{\begin{tabular}[c]{@{}l@{}}AlderLake-S\cite{alderlake-micro}\\ (i7-12700F)\end{tabular}} \\
\multicolumn{1}{l|}{Peak Freq.}                                                                  & 3.2GHz                                                                 & 3.2GHz                                                                 & 2.8GHz                                                                    & 4.9GHz                                                                     \\
\multicolumn{1}{l|}{Tech.}                                                                    & 14nm                                                                   & 5nm                                                                    & 7nm                                                                       & 10nm                                                                       \\
\multicolumn{1}{l|}{Area}                                                                     & 1.11$mm^2$                                                             & 2.53$mm^2$                                                             & 1.23$mm^2$                                                                & 7.30$mm^2$                                                                 \\
\multicolumn{1}{l|}{@14nm}                                                                    & 1.11$mm^2$                                                             & 22.55$mm^2$                                                            & 3.61$mm^2$                                                                & 22.63$mm^2$                                                                \\
\multicolumn{1}{l|}{IPC}                                                                      & 1.3                                                                    & 3.79                                                                   & 2.07                                                                      & 2.83                                                                       \\
\multicolumn{1}{l|}{\begin{tabular}[c]{@{}l@{}}\textbf{Normalized}\\\textbf{Throughput}\end{tabular}}                                                            & \textbf{1}                                                             & \textbf{2.92}                                                          & \textbf{1.27}                                                             & \textbf{3.35}                                                              \\ \hline
\multicolumn{5}{c}{\textit{FireGuard Elements}}                                                                                                                                                                                                                                                                                                                                                          \\ \hline
\multicolumn{1}{l|}{Filter Width}                                                             & 4-way                                                                  & 8-way                                                                  & 4-way                                                                     & 6-way                                                                      \\
\multicolumn{1}{l|}{\#$\upmu$cores}                                                             & 4                                                                      & 12                                                                     & 5                                                                         & 13                                                                         \\
\multicolumn{1}{l|}{\textbf{\begin{tabular}[c]{@{}l@{}}Overhead \\ (\% / Core)\end{tabular}}} & \textbf{\begin{tabular}[c]{@{}l@{}}0.29$mm^2$\\ (25.9\%)\end{tabular}} & \textbf{\begin{tabular}[c]{@{}l@{}}0.81$mm^2$\\ (3.6\%)\end{tabular}}  & \textbf{\begin{tabular}[c]{@{}l@{}}0.35$mm^2$\\ (9.6\%)\end{tabular}}     & \textbf{\begin{tabular}[c]{@{}l@{}}0.85$mm^2$\\ (3.8\%)\end{tabular}}      \\ \hline
\multicolumn{5}{c}{\textit{An Independent Kernel for All Cores}}                                                                                                                                                                                                                                                                                                                                          \\ \hline
\multicolumn{1}{l|}{\textbf{\begin{tabular}[c]{@{}l@{}}Overhead \\ (\% / SoC)\end{tabular}}}  & \textbf{\begin{tabular}[c]{@{}l@{}}0.29$mm^2$\\ (9.86\%)\end{tabular}} & \textbf{\begin{tabular}[c]{@{}l@{}}6.10$mm^2$\\ (0.47\%)\end{tabular}} & \textbf{\begin{tabular}[c]{@{}l@{}}1.23$mm^2$\\ (0.57\%)\end{tabular}}    & \textbf{\begin{tabular}[c]{@{}l@{}}6.67$mm^2$\\ (0.99\%)\end{tabular}}     \\ \hline
\toprule
\end{tabular}}
\caption{Feasibility of \marchname\ in commercial SoCs.} 
\label{table:fa}
\end{table}

\subsection{Feasibility Analysis}
\label{sc:FA}

\label{FA:Processors}
% \hugo{Add some descriptions regarding the methods used in this section.\\}

% \Cref{table:fa}'s upper portion compares throughput and area between BOOM and performance cores in commercial SoCs~\cite{m1pro-overview,a76-arch,alderlake-spec}.
% The middle portion estimates the overhead to support a kernel for the cores, while attaining with the performance reported in \cref{sbsc:PerformanceOverhead}. Since these SoCs are fabricated using different technologies and operate under varying frequencies, we normalize the area using 14nm technology~\cite{fabrication} and the throughput (Insts/Cycle) using IPC $\times$ peak frequency, measured through Parsec execution.

We compare \emph{performance-area trade-offs} between BOOM and the OoO processors in commercial SoCs~\cite{m1pro-overview,m1pro-micro,a76-overview, a76-arch,rotem2022intel,alderlake-spec,rotem2022intel,rotem2022intel}, and examine the overhead associated while constructing \marchname\ upon these processors and SoCs.
To do so, we first estimate the core sizes from die shots of each system~\cite{m1pro-overview, alderlake-micro, a76-overview}, then normalize the area using 14 $nm$ technology based on the density difference~\cite{fabrication}.
We then scale up the number of $\mu$cores to match the increased IPC and clock relative to BOOM (calculated using IPC $\times$ peak frequency), where the IPC is measured through single-thread Parsec execution.
% Such scaling is realistic since $\mu$core workloads display extreme parallelism.

\parlabel{CPU-level overhead.}
Table~\ref{table:fa}'s upper portion examines throughput and area between BOOM and performance cores in the SoCs;
the middle portion estimates the overhead to support a kernel, while attaining the performance in \cref{sbsc:PerformanceOverhead}.

Modern OoO cores are significantly larger than BOOM.
To achieve a linear increase in processor throughput, a superlinear increase in hardware overhead is required. 
FireStorm is 2.92$\times$ faster than BOOM, but each core consumes 20.3$\times$ the area.
While building \marchname\ with a bigger core would require extra $\upmu$cores to keep up with the execution, the increase is only linear.
Building \marchname\ upon FireStorm, Cortex-A76, and AlderLake-S requires 12, 5, and 13 $\upmu$cores respectively, giving 3.6\%, 9.6\%, and 3.8\% overhead/core.

\parlabel{SoC-level overhead.}
\label{FA:SoCs}
To ensure global security, \marchname\ elements must be equipped for all processors in the SoC.
Hence, we apply the same analysis on all core types in the SoCs and report the overhead. % while deploying an independent kernel for all cores.

Table~\ref{table:fa}'s bottom portion indicates that integrating a kernel in M1-Pro, Kirin-980, and i7-12700F leads to an overhead of less than 1\%. 
This negligible impact makes implementing several kernels per-core practical. 
Moreover, the energy overhead would be even lower, since the majority of \marchname\ operates within a low-frequency domain.

\section{Conclusion}
\label{sc:Conclusion}

We have presented \marchname{}, a microarchitecture for fine-grained instruction analysis.
To make the design practical for deployment, we presented a buffer-free data forwarding channel, a superscalar event filter and a broadcast-free mapper.
Feasibility analysis shows \marchname\ can be integrated\ into modern SoCs with less than 1\% increase in area.
In summary, comprehensive in-core analysis is practical and efficient to build into real cores and SoCs.

\section{Acknowledgement}
We appreciate the anonymous reviewers for their helpful feedback. 
This work is supported by the National Key Research and Development Program (Grant No. 2024YFB4405600), the National Natural Science Foundation of China (Grant No. 62472086, 62204036) and the Basic Research Program of Jiangsu (Grants No. BK20243042).

\newpage

\linespread{.98}
\bibliographystyle{IEEEtran}
\bibliography{refs}

\end{document}